\documentclass[sn-mathphys]{sn-jnl}

\usepackage{graphicx}
\usepackage{amsmath}
\usepackage{amsfonts}
\usepackage{amssymb}
\usepackage{bm}
 \usepackage{mathtools}
\usepackage{xcolor}

\jyear{2021}%

\theoremstyle{thmstyleone}%
\newtheorem{theorem}{Theorem}
\theoremstyle{thmstyletwo}%

\theoremstyle{thmstylethree}%
\newtheorem{definition}{Definition}%

\newcommand{\zeri}{{\scriptsize\stackrel{n-h}{\overbrace{0 \cdots 0}}}}

\newcommand{\cc}{{\boldsymbol \xi}}

\newcommand{\xx}{\mathbf{x}}

\newcommand{\II}{\mathcal{I}}

\newcommand{\da}{\, \varphi(\mathcal{I}) \, d\mathcal{I} d {\boldsymbol \xi}}
\newcommand{\inta}{\int_{\mathbb{R}^{3}}\int_0^{+\infty}}

\begin{document}

	\title[Relativistic Kinetic Theory and RET of  Polyatomic Gases]{Relativistic Kinetic Theory of  Polyatomic Gases: 
		Classical Limit of  a   New Hierarchy of Moments  and Qualitative Analysis}


\author[1]{\fnm{Takashi} \sur{Arima}}\email{arima@tomakomai-ct.ac.jp}
\equalcont{These authors contributed equally to this work.}

\author[2]{\fnm{Maria Cristina} \sur{Carrisi}}\email{mariacri.carrisi@unica.it}
\equalcont{These authors contributed equally to this work.}

\author[2]{\fnm{Sebastiano} \sur{Pennisi}}\email{spennisi@unica.it}
\equalcont{These authors contributed equally to this work.}

\author*[3]{\fnm{Tommaso} \sur{Ruggeri}}\email{tommaso.ruggeri@unibo.it}
\equalcont{These authors contributed equally to this work.}

\affil[1]{\orgdiv{Department of Engineering for Innovation}, \orgname{National Institute of	 Technology, Tomakomai College},  \orgaddress{  \city{Tomakomai},   \country{Japan}}}

\affil[2]{\orgdiv{Department of  Mathematics and Informatics}, \orgname{University of Cagliari},  \orgaddress{  \city{Cagliari},   \country{Italy}}}

\affil*[3]{\orgdiv{Department of Mathematics and Alma Mater Research Center on Applied Mathematics AM$^2$}, \orgname{	University of Bologna},   \orgaddress{ \city{Bologna},  \country{Italy}}}


\abstract{A relativistic version of the Kinetic  Theory for polyatomic gas is considered and a new hierarchy of moments that takes into account the total energy composed by the rest energy and the energy of the molecular internal modes is presented. In the first part, we prove via classical limit that the truncated system of moments dictates a precise hierarchy of moments in the classical framework. In the second part, we consider the particular physical case of fifteen moments closed via Maximum Entropy Principle in a neighborhood of equilibrium state. We prove that this   symmetric hyperbolic  system satisfies all the  general assumption of some theorems that guarantee the global existence of smooth solutions for initial data sufficiently small.}




\maketitle

\section{Introduction}
The kinetic theory offers an excellent  mathematic model for  rarefied gases. The celebrated Boltzmann equation \footnote{As usual repeated indices indicated omitted sum symbol.}
\begin{equation} \label{eq:Boltzmann}
 \partial_t f^C + \xi^i\, \partial_i  f^C= Q^C,
\end{equation}
is widely used in many applications and is still now a challenge for its difficult mathematical questions.
The state of the gas is  described by the distribution function $f^C(\xx, t, \cc)$, 
being respectively $\xx \equiv(x^i),\cc\equiv(\xi^i),t$ the space coordinates, the microscopic velocity and the time.  $Q^C$ denotes the collisional term and $\partial_t = \partial /\partial t$, $\,\, \partial_i = \partial/\partial x_i$, ($i=1,2,3$). There are many results on Boltzmann equation, in particular we quote for the mathematical treatment the books of  Cercignani \cite{Cer1,Cer2} who was one of the world leaders that gave fundamental papers on this subject.

The relativistic counterpart of Boltzmann equation is   the Boltzmann-Chernikov  equation
\cite{BGK,Synge,KC}:
\begin{equation}\label{BoltzR}
p^\alpha \partial_\alpha f = Q,
\end{equation}
in which the relativistic distribution function  $f$ depends on  $(x^\alpha,  p^\beta)$, where $x^\alpha$ are the space-time coordinates, $p^\alpha$ is the four-momentum, $\partial_{\alpha} = \partial/\partial x^\alpha$,    $Q$ is the collisional term and $\alpha, \beta =0,1,2,3$. 

Formally the relativistic equation converges to the classical one if we take into account the following expressions  (see for example \cite{Annals})
\begin{align}\label{limitini}
\begin{split}
&x^0= c t, \quad  p^0 = m \, \Gamma \, c \, , \quad  p^i = m \, \Gamma \, \xi^i \, , \quad   \Gamma = \left( 1- \, \frac{\xi^2}{c^2} \right)^{- \frac{1}{2}} ,  \\
&\lim_{c \rightarrow \, + \infty} f= \frac{1}{m^3} \, f^C \, , \quad \lim_{c \rightarrow \, + \infty} Q= \frac{1}{m^2} \, Q^C ,
\end{split}
\end{align}
where $c$ denotes the light velocity, $m$ is the particle mass in the rest frame and $\Gamma$ is the Lorentz factor.

The weak point of the Boltzmann equation both in classical and relativistic regimes is that their  validity holds only for monatomic gas even if the classical kinetic theory was used in fields very far from gas dynamics like in biological phenomena, socio-economic systems, models of swarming, and many other fields (see, for example, \cite{Preziosi1,Tosin,Carillo} and references therein).

A more realistic case which is important for applications is the kinetic theory of polyatomic gas. In the classical framework was proposed based on two different approaches: 

-  the description of the  internal structure   of a polyatomic gas   is taken into account by a large number of discrete energy states, so that the gas might be considered as a sort of mixture of monatomic components, which interact by binary collisions with conservation of total energies, but with possible exchange of energy between its kinetic and internal (excitation) forms. The model can be used also in a reactive frame, even in the presence of its self--consistent radiation field \cite{GroppiSpiga}.

\vskip 6pt

- Another  approach in the development
of the theory of rarefied polyatomic gases was made by Borgnakke and Larsen \cite{Borgnakke-1975}.
The distribution function is assumed to depend on an additional continuous variable  $\cal{I}$
representing the energy of the internal modes of a molecule in order to take into account the exchange
of energy (other than translational one) in binary collisions. This model was initially used
for Monte Carlo simulations of polyatomic gases, and later it has been applied to the derivation
of the generalized Boltzmann equation by Bourgat,  Desvillettes,   Le Tallec   and Perthame   \cite{Bourgat-1994}.
In this case the Boltzmann equation \eqref{eq:Boltzmann} has the same form but the
distribution function $f^C(\xx, t, \cc,\cal{I})$ is
defined on the extended domain ${R}^{3} \times  [0,\infty) 
\times  {R}^{3} \times [0,\infty)$ 
and  the collision integral   takes into account
the influence of the internal degrees of freedom through the collisional 
cross section. The case of  non polytropic gases in which the internal energy is a non-linear function of the temperature  was considered by Ruggeri and coworkers in a series of papers \cite{rendmat,RuggeriSpiga,Ruggeri-2020RdM}. A more refined case in which the internal mode is divided into the rotational and vibrational modes was presented  by Arima, Ruggeri and Sugiyama \cite{ET7,ET15a}. Concerning the production terms it was used   a BGK model \cite{Ruggeri-2020RdM} or  an extended one with two or more relaxation times \cite{Struch1,Struch2,ET7,ET15a}.

In the relativistic framework, Pennisi and Ruggeri \cite{Annals} used a similar technique, and they postulate  the same Boltzmann-Chernikov equation \eqref{BoltzR} but with a distribution function $f(x^\alpha,p^\alpha,\cal{I})$ depending on the   microscopic energy due to the internal mode. The  same authors construct a new BGK  model both for monatomic and for polyatomic gas in \cite{BGKPR}. The existence and asymptotic behavior of classical solutions for this model when the initial data is sufficiently close to a global equilibrium was the subject of the paper \cite{Koreani}.

Both in the classical and relativistic theory, we can associate macroscopic quantities, called moments, which satisfy an infinite set of balance laws. The choice of the moments is a controversial question in particular in the polyatomic gas.  

The closure of moments when the number is finite is the starting point of modern Rational Extended Thermodynamics (RET). The aim of this paper is to discuss first the classical limit of a new hierarchy or relativistic moments and in the particular case of a RET  with 15 moments  to discuss the qualitative analysis of the solutions. In particular, we prove that these systems, both in relativistic and classical cases, satisfy the conditions of the well-known theorems for the existence of the global smooth solutions for initial data sufficiently small.

In the present paper, first we discuss  the relativistic Boltzmann-Chernikov equation for polyatomic gases and after in Sec. 2, we present a brief review on the possible  physical moments that take  into account the total energy composed by the rest energy and the energy of molecular internal states. Then  in Sec. 3, we will study the non-relativistic limit of the system of balance equations for any number of moments.   As a particularly interesting case, we summarize in Sec. 4  the result of the  RET  theory with 15 moments. In Sec. 5, we show that for the relativistic RET  with 15 moments and its classical limit hold the theorems of the existence of the global smooth solutions under given sufficiently small initial data.

\section{Moments associated to the Kinetic Equation}

In the classical case and for monatomic gases  the moments are:
\begin{equation} 
F_{k_1 k_2 \cdots k_j}= m\int_{\mathbb{R}^{3}}{f^C \xi_{k_1} \xi_{k_2} \cdots \xi_{k_j} d{\bm \xi} },\qquad (j=0,1,\dots ), \label{prima}
\end{equation}
$k_1,k_2,\dots =1,2,3$ and
 by convention when $j=0$ we have 
\begin{equation*} 
F = m \int_{\mathbb{R}^{3}}{f   d{\bm \xi} }.
\end{equation*}

Due to the Boltzmann equation \eqref{eq:Boltzmann}, the moments satisfy an infinite hierarchy of balance laws in
which the flux in one equation becomes the density in the next one:
\begin{equation} \label{F-Gerarchia}
\partial_t F_A + \partial_i F_{iA}= P_{A}, \qquad (A=0,1,\dots) 
\end{equation}
where we introduce the following  multi-index notation:
\begin{equation} \label{F-Mom}
F_A = F_{i_1 i_2\dots i_A}, \quad    F_{iA} = F_{ii_1 i_2\dots i_A}, \quad   P_A = P_{i_1i_2\dots i_A},
\end{equation}
with
\begin{equation}\label{ggg}
P_{k_1 k_2 \cdots k_j}= m \int_{\mathbb{R}^3}{Q \xi_{k_1} \xi_{k_2} \cdots \xi_{k_j} d\boldsymbol{\xi}}.
\end{equation}

The relativistic counterpart of the moment  equations for monatomic gas are 
\begin{equation}\label{MRel}
\partial_\alpha A^{\alpha \alpha_1 \cdots \alpha_n  } =  I^{  \alpha_1 \cdots \alpha_n   }
\quad \mbox{with} \quad n=0 \, , \,\cdots \, , \,  
\end{equation}
with  
\begin{align*}
\begin{split}
& A^{\alpha \alpha_1 \cdots \alpha_n  } = \frac{c}{m^{n-1}} \int_{\mathbb{R}^{3}}
f  \,  p^\alpha p^{\alpha_1} \cdots p^{\alpha_n}  \, \, d \boldsymbol{P}, \\
& I^{\alpha_1 \cdots \alpha_n  } = \frac{c}{m^{n-1}} \int_{\mathbb{R}^{3}}
Q  \,   p^{\alpha_1} \cdots p^{\alpha_n}  \, \, d \boldsymbol{P} ,
\end{split}
\end{align*}
where  the Greek indices run from $0$ to $4$,   and 
\begin{equation*}
d \boldsymbol{P} =  \frac{dp^1 \, dp^2 \,
	dp^3}{p^0} .
\end{equation*}
If we truncate the moments \eqref{MRel} until the index $N$, i.e. $n=0,1,2,\dots,N$, it was proved in  \cite{JSP} the following theorem:
\begin{theorem}[Pennisi-Ruggeri \cite{JSP}]\label{CLRM}
- 	For a prescribed truncation index $N$, for any   integer
	$0\leq s \leq N$ and for  multi-index $0\leq B \leq N-s$, the  relativistic moment system for a monatomic  gas   \eqref{MRel} with $n = 0,1,\dots,N$
	converges, when  $c\rightarrow \infty$, to the following classical moments system:
	\begin{align} \label{HSAM}
	\begin{split}
	&    \partial_t  F_{B}+ \partial_i  F _{iB} = P _{B}, \quad \text{if} \quad s=0, \quad 0\leq B\leq N, \quad  \text{and}  \\
	&      \partial_t  F_{j_1 j_1\dots j_s j_s i_1 i_2 \dots i_{N-s}}+ \partial_i  F _{i j_1j_1 \dots j_s j_s i_1 i_2 \dots i_{N-s}} = P _{j_1 j_1\dots j_s j_s i_1 i_2 \dots i_{N-s}},
	\end{split}
	\end{align}
	 with  $1\leq s \leq N$.
The   moments $F$'s are given by \eqref{prima} and the productions $P$'s are  given by \eqref{ggg}.
	In particular, for $s=0$,  we have the  $ F$'s  moments with all free indexes  until index of truncation $N$ and for $1\leq s \leq N$ there is a single block of $F$'s  moments with increasing number of pairs of contracted indexes. The truncated tensorial index in \eqref{HSAM} is $\bar{N} = 2 N$.
\end{theorem}
This theorem solved the old  problem  how to choose in an optimal way the moments in the classical case. In fact, there are several degrees of freedom depending on how many indices are saturated in the truncated tensors. For example, in the Grad system for which in \eqref{F-Gerarchia} $\bar{N}=3$  instead of taking all free indices, it was considered  two indexes saturated  in the triple tensor:  $(F,F_i,F_{ij},F_{kki})$.

We remark that, for $N=1$, the system \eqref{MRel} is the Euler relativistic fluid and the classical limit is the Euler classical fluid with moments $(F,F_i,F_{ll})$ of which balance laws correspond to the mass, momentum and energy conservations. While, for $N=2$, the relativistic system \eqref{MRel} is the one proposed by Liu, M\"uller and Ruggeri \cite{LMR} and the classical limit converges to the $14$ moments proposed by Kremer \cite{Kremer}: $(F,F_i,F_{ij},F_{lli},F_{kkjj})$,  instead of the Grad system. The Grad $13$ moments  don't correspond to any  classical limit of a relativistic theory!

\bigskip

How to construct moments in the polyatomic case was an open problem. Starting from the equilibrium  case of $5$ moments proposed by Bourgat et al. \cite{Bourgat-1994} a double hierarchy of moments  was proposed first at macroscopic level with $14$ field by Arima et al. \cite{dense} and successively at kinetic level in the papers \cite{Pavic-2013,AMR}, (see for more details \cite{newbook}):
\begin{align*}
\begin{split}
&  F_{i_1 \ldots i_j} = m \inta  f^C \, \xi_{i_1}\cdots \xi_{i_j}  \,  \da ,    \\ 
&   {G_{lli_1 \ldots i_k}}  =  2\inta f^C \, \left(\frac{m \xi^2}{2} + \II\right) \xi_{i_1}\cdots \xi_{i_k}  \,  \da .
\end{split}
\end{align*}
Here $\varphi(\II)$ is the state density corresponding to $\II$, i.e., $\varphi(\II) d\II$ represents the number of internal state between $\II$ and $\II + d\II$. As (for $k=0$) $G_{ll}$ is the energy, except for a factor $2$, we have that the $F's$ are the usual \emph{momentum-like} moments  and the $G's$ are \emph{energy-like} moments.

From the Boltzmann equation \eqref{eq:Boltzmann}, we obtain a binary hierarchy of balance equations   so called $(F,G)$-hierarchies:
\begin{align}\label{momentiniFG}
\begin{split}
& \partial_t F_{k_1 k_2 \dots k_n}  \hspace{0.3cm}+\partial_i F_{k_1 k_2 \dots k_n i} \,\, \,\, = P_{k_1k_2 \dots k_n}, \qquad \quad \,\,\,\, n=0,1,\dots , \bar{N}, \\
& \partial_t G_{ll k_1 k_2 \dots k_m}  +  \partial_i  G_{ll k_1 k_2 \dots k_m i} = Q_{ll k_1k_2 \dots k_m}, \qquad \,\,\, m=0,1,\dots , \bar{M}.
\end{split}
\end{align} 
From the requirement of the Galilean invariance and the physically reasonable solutions, it is shown that $\bar{M}=\bar{N}-1$ \cite{AMR}. The case with $\bar{N}=1$ corresponds to the Euler system, and the one with $\bar{N}=2$ corresponds to RET with $14$ moments \cite{dense,Pavic-2013}.

 Pennisi and Ruggeri first in \cite{Annals}  and then in \cite{JSP} proved that the relativistic theory of moments for polyatomic case contains  in the classical limit  the $(F,G)$-hierarchy if we consider a system \eqref{MRel} but with the following moments:
\begin{align} 
\begin{split}
& A^{\alpha \alpha_1 \cdots \alpha_n  } = \frac{1}{m^n c} \inta  f  \,  p^\alpha p^{\alpha_1} \cdots p^{\alpha_n}  \, \left(mc^2 +  n \II \right)\, 
\phi(\mathcal{I}) \, d \mathcal{I} \, d \boldsymbol{P}  \, , \\
& I^{\alpha_1 \cdots \alpha_n  } = \frac{1}{m^{n}c} \inta Q  \,  p^{\alpha_1} \cdots p^{\alpha_n}  \, \left( mc^2 +  n\II \right)\, 
\phi(\mathcal{I}) \, d \mathcal{I} \, d \boldsymbol{P}, \\
\end{split}
\label{relRETold}
\end{align}
where the distribution function $f(x^\alpha, p^\beta,\mathcal{I})$ depends on the extra energy variable $\mathcal{I}$, similar to the classical one.

%

Pennisi in \cite{Pennisi-2021} noticed first the unphysical situation in which, instead to have  the full energy at molecular level, i.e., $mc^2 +\II$ , we have in \eqref{relRETold}$_2$ the term $mc^2 + n\II$ but he  observed  that  $(mc^2)^{n-1}(mc^2 +n \II)$   are the first two terms of  the Newton binomial formula of  $(mc^2 +\mathcal{I})^n/ (mc^2)^{n-1}$.
Therefore he proposed in \cite{Pennisi-2021} to modify, in the relativistic case, the definition of the moments by using the substitution (see also \cite{entropy}):
\begin{equation*}
(mc^2)^{n-1}\left( mc^2 + n \II \right) \qquad \text{with \quad } \left( mc^2 + \II \right)^n,
\end{equation*}
i.e., instead of \eqref{relRETold}, the following moments were proposed:
\begin{align} \label{relRET}
& A^{\alpha \alpha_1 \cdots \alpha_n  } = \left(\frac{1}{mc}\right)^{2n-1} \inta  f  \,  p^\alpha p^{\alpha_1} \cdots p^{\alpha_n}  \, \left( mc^2 +  \II \right)^n\, 
\phi(\mathcal{I}) \, d \mathcal{I} \, d \boldsymbol{P}  \, , \nonumber \\
& \\
& I^{\alpha_1 \cdots \alpha_n  } = \left(\frac{1}{mc}\right)^{2n-1} \inta Q  \,  p^{\alpha_1} \cdots p^{\alpha_n}  \, \left( mc^2 +  \II \right)^n\, 
\phi(\mathcal{I}) \, d \mathcal{I} \, d \boldsymbol{P}. \nonumber
\end{align}
In the next section we determine  what is the classical limit of the truncated system \eqref{MRel} with $n=0,1,\dots,N$ and  moments given by \eqref{relRET}.
\section{The non relativistic limit}
In this section we prove the following
\begin{theorem}\label{th2}
	For a prescribed truncation integer index $N$  and $0\leq s \leq N$, the  relativistic moment system for polyatomic gases  \eqref{MRel}  (with $n=0,1,\dots,N$) and  \eqref{relRET}, converges when  $c\rightarrow +\infty$ to the following $N+1$ hierarchies of classical moments:
	\begin{align}\label{2.3}
\begin{split}
	 \partial_t H_{s}^{i_1 \cdots i_h} +&  \partial_i \, H_{s}^{i \, i_1 \cdots i_h} =  J_s^{ i_1 \cdots i_h  } \\
	& \mbox{with } s=0 \, , \, \cdots \, , \,  N   \mbox{ and }  h=0 \, , \,\cdots \, , \,  N - s  
\end{split}
	\end{align} 
	where 
	\begin{align}\label{2.4}
	&    H_{s}^{i_1 \cdots i_h} = m  \int_{\mathbb{R}^{3}}
	\int_0^{+\infty} f^C \, \xi^{i_1} \cdots \xi^{i_h} 
	\,  \left( \frac{2 \, \mathcal{I}}{m} \, + \xi^2 \right)^{s} \,  \phi(\mathcal{I}) \, d \, \mathcal{I} \, d{\bm \xi}   \, , 	\nonumber \\
	&     H_{s}^{i \, i_1 \cdots i_h} = m  \int_{\mathbb{R}^{3}}
	\int_0^{+\infty} f^C \,\xi^i  \xi^{i_1} \cdots \xi^{i_h} 
	\,  \left( \frac{2 \, \mathcal{I}}{m} \, + \xi^2 \right)^{s} \,  \phi(\mathcal{I}) \, d \, \mathcal{I} \, d{\bm \xi}   \, ,   \\
	& J_s^{ i_1 \cdots i_h  }    = m  \int_{\mathbb{R}^{3}}
	\int_0^{+\infty} Q^C \, \xi^{i_1} \cdots \xi^{i_h} 
	\,  \left( \frac{2 \, \mathcal{I}}{m} \, + \xi^2 \right)^{s} \,  \phi(\mathcal{I}) \, d \, \mathcal{I} \, d{\bm \xi}   \, , \nonumber
	\end{align}
	$f^C$ and $Q^C$ are the classical limits of $f$ and $Q$ respectively.
	In particular, for $s=0$  we have the  momentum-like   block of equations \eqref{momentiniFG}$_1$, for  $s=1$ the  energy-like  block  \eqref{momentiniFG}$_2$ and for $2\leq s \leq N$ there are new blocks never considered before in the literature.
\end{theorem}

\textbf{Proof}: Let us write our equations in 3-dimensional form. Taking into account that $x^ 0=c \, t$ and $\partial_0   = {1}/{c}  \,  \partial_t $, they become
\begin{equation*}
\frac{1}{c} \, \partial_t A^{0 \alpha_1 \cdots \alpha_n  } + \partial_i A^{i \alpha_1 \cdots \alpha_n  }=  I^{  \alpha_1 \cdots \alpha_n   } \, ,
\end{equation*}
or
\begin{align}\label{6a}
\partial_t A^{0 \, {\zeri} \,  i_1 \cdots i_h  } + \partial_i  \left(c \, A^{i \, \zeri\,  i_1 \cdots i_h  } \right)=  c \, I^{\zeri\,  i_1 \cdots i_h  } \\
\mbox{with} \quad h=0 \, , \, \cdots \, , \,  n \, , \quad \mbox{and} \quad  n=0 \, , \,\cdots \, , \,  N \, . \nonumber
\end{align}
Here $\stackrel{n-h}{\overbrace{0 \cdots 0}}$ represents a set of $n-h$ zeros. From \eqref{limitini} and \eqref{relRET},  we have  
\begin{align*}
A^{i \, \zeri\,  i_1 \cdots i_h  }=  m \,  \int_{\mathbb{R}^{3}}
\int_0^{+\infty} f   \, c^{n-h} \Gamma^{n+5} \xi^{i} \xi^{i_1} \cdots \xi^{i_h}  \, \left( 1 + \frac{\mathcal{I}}{m \, c^2} \right)^n \, 
\phi(\mathcal{I}) \,d \,  \mathcal{I}  \,  d{\bm \xi}  \,  \, , 
\end{align*}
and
\begin{align*}
A^{0 \, \zeri\,  i_1 \cdots i_h  }=  m \,  \int_{\mathbb{R}^{3}}
\int_0^{+\infty} f  \, c^{n-h+1} \Gamma^{n+5} \xi^{i_1} \cdots \xi^{i_h}  \, \left( 1 + \frac{\mathcal{I}}{m \, c^2} \right)^n \, 
\phi(\mathcal{I}) \,d \,  \mathcal{I}  \,  d{\bm \xi}  \,  \, .
\end{align*}
Eq. \eqref{6a} divided by $c^{n-h+1}$ becomes
\begin{align}\label{2.1}
& \partial_t \tilde{A}_n^{ i_1 \cdots i_h  } + \partial_i  \, \tilde{A}_n^{i \, i_1 \cdots i_h  } =  \tilde{I}_n^{ i_1 \cdots i_h  } \\
& \mbox{for} \quad h=0 \, , \, \cdots \, , \,  n \, , \quad \mbox{and} \quad  n=0 \, , \,\cdots \, , \,  N \, , \nonumber
\end{align} 
with 
\begin{eqnarray}\label{2.2}
&    \tilde{A}_n^{ i_1 \cdots i_h  }=  m^4 \,  \inta  {f   \,  \Gamma^{n+5} \xi^{i_1} \cdots \xi^{i_h}  \, \left( 1 +  \frac{\mathcal{I}}{m \, c^2} \right)^n \, 
\phi(\mathcal{I}) \,d \,  \mathcal{I}  \,  d{\bm \xi}  \, } \, , \\
& \nonumber \\
&    \tilde{A}_n^{ i i_1 \cdots i_h  }=  m^4 \,  \int_{\mathbb{R}^{3}}
\int_0^{+\infty} f   \,  \Gamma^{n+5} \xi^i \xi^{i_1} \cdots \xi^{i_h}  \, \left( 1 +  \frac{\mathcal{I}}{m \, c^2} \right)^n \, 
\phi(\mathcal{I}) \,d \,  \mathcal{I}  \,  d{\bm \xi}  \,  \, , \nonumber  \\
& \tilde{I}_n^{ i_1 \cdots i_h  }    = \frac{1}{c^{n-h}} \,  I^{ \,  i_1 \cdots i_h  } = 
m^3 \,  \int_{\mathbb{R}^{3}}
\int_0^{+\infty} f   \,  \Gamma^{n+4} \xi^i \xi^{i_1} \cdots \xi^{i_h}  \, \left( 1 +  \frac{\mathcal{I}}{m \, c^2} \right)^n \, 
\phi(\mathcal{I}) \,d \,  \mathcal{I}  \,  d{\bm \xi}  \,  \nonumber \, . 
\end{eqnarray}
We can see that the equations \eqref{2.1} with different n but the same value of $h$ have the same non relativistic limit, so that the number of independent equations is reduced. In order to preserve the number of independent  equations, for every fixed value of $h$, we define a new tensor as a linear combination of the $\tilde{A}_n^{ i_1 \cdots i_h  }$ from \eqref{2.2}$_1$: 
\begin{align}\label{3.2}
& H_{\mbox{rel} , N-n}^{i_1 \cdots i_h} \stackrel{\mbox{def}}{=}\left( 2 \, c^2 \right)^{N-n} \, \sum_{r=0}^{N-n} \, \begin{pmatrix}
\, N-n \,  \nonumber\\
\, r \,
\end{pmatrix} \, (-1)^{N-n-r}   \tilde{A}_{r+n}^{i_1 \cdots i_h} = \nonumber \\
& m^4 \left( 2 \, c^2 \right)^{N-n} \int_{\mathbb{R}^{3}}
\int_0^{+\infty} f \, \xi^{i_1} \cdots \xi^{i_h} \nonumber \\
& \hspace{1cm} \sum_{r=0}^{N-n} \begin{pmatrix}
\, N-n \, \\
\, r \,
\end{pmatrix}
\, (-1)^{N-n-r} \, \Gamma^{r+n+5}  \,  \left( 1 +  \frac{\mathcal{I}}{m \, c^2} \right)^{r+n} 
\phi(\mathcal{I}) \, d \, \mathcal{I} \, d{\bm \xi}  \nonumber \\
& = m^4  \int_{\mathbb{R}^{3}}
\int_0^{+\infty} f \, \xi^{i_1} \cdots \xi^{i_h} \, \Gamma^{n+5}
\,  \left( 1 +  \frac{\mathcal{I}}{m \, c^2} \right)^{n} \nonumber \\
&\hspace{2cm}
\left\{ \underline{2 \, c^2 \, \left[ \Gamma \left( 1 +  \frac{\mathcal{I}}{m \, c^2} \right) \, - \, 1 \right]} \right\}^{N-n}
\phi(\mathcal{I}) \, d \, \mathcal{I} \, d{\bm \xi}. \nonumber
\end{align}
The non relativistic limit of the underlined part of the above expression is $2\mathcal{I} /m\, + \xi^2$, so that 
\begin{align*}
\begin{split}
& \lim_{c \, \rightarrow \, + \infty} H_{\mbox{rel} , N-n}^{i_1 \cdots i_h}  = H_{N-n}^{i_1 \cdots i_h}  \quad \mbox{with} \\
& H_{N-n}^{i_1 \cdots i_h} = m  \int_{\mathbb{R}^{3}}
\int_0^{+\infty} f^C \, \xi^{i_1} \cdots \xi^{i_h} 
\,  \left( \frac{2 \, \mathcal{I}}{m} \, + \xi^2 \right)^{N-n} \,  \phi(\mathcal{I}) \, d \, \mathcal{I} \, d{\bm \xi}   \, ,
\end{split}
\end{align*}
for $h=0 \, , \, \cdots \, , \,  n \, , \quad \mbox{and} \quad  n=0 \, , \,\cdots \, , \,  N$.  This set of indexes can be expressed also by the conditions $0 \leq h \leq   N \, , \quad  0 \leq n \leq   N$ and $h \leq n$. Now we can change index according to the law $N-n=s$ so that 
\begin{align*}
H_{s}^{i_1 \cdots i_h} = m  \int_{\mathbb{R}^{3}}
\int_0^{+\infty} f^C \, \xi^{i_1} \cdots \xi^{i_h} 
\,  \left( \frac{2 \, \mathcal{I}}{m} \, + \xi^2 \right)^{s} \,  \phi(\mathcal{I}) \, d \, \mathcal{I} \, d{\bm \xi}   \, ,
\end{align*}
and the above set of indexes transforms in $0 \leq h \leq   N \, , \quad  0 \leq s \leq   N$ and $h \leq N-s$ or, equivalently, for $s=0 \, , \, \cdots \, , \,  N \, , \quad \mbox{and} \quad  h=0 \, , \,\cdots \, , \,  N-s$. Eq. $\eqref{2.4}_1$ is proved. \\
The same passages can be followed starting from \eqref{2.2}$_2$, \eqref{2.2}$_3$.
Finally, starting from \eqref{2.1} we can prove our theorem, showing that the non relativistic limit is \eqref{2.3}.  

\smallskip

We observe that also in the classical limit now appears in the moments \eqref{2.2}  the full energy given by the sum of kinetic energy  plus the energy of internal modes: $m \xi^2/2+\cal{I}$.

\subsection{Particular cases}
As example we consider the case $N=1$. The relativistic moments \eqref{MRel} with $n=0,1$ reduces now to
\begin{equation}\label{3.2}
\partial_{\alpha} A^\alpha = 0, \qquad \partial_{\alpha} A^{\alpha  \beta} = 0
\end{equation}
that corresponds to the Euler relativistic polyatomic gas. The corresponding limit according with the Theorem \ref{th2}  is:
\begin{align*}
& s=0 \,\, \| \quad \partial_t H_0^0 +\partial_i H_0^i = 0,\, \quad \leftrightarrow \quad  \partial_t F +\partial_i F^i= 0, \quad \,\, \text{(mass)} \nonumber\\
&\quad \quad \quad \,\,\,\,\,\, \partial_t H_0^j +\partial_i H_0^{ij} = 0, \quad \leftrightarrow \quad  \partial_t F^j +\partial_i F^{ij}= 0, \quad \text{(momentum)} \\
& s=1 \,\, \| \quad \partial_t H_1^0 +\partial_i H_1^i = 0,\, \quad \leftrightarrow \quad  \partial_t G_{ll} +\partial_i G^{lli}= 0, \quad \text{(energy)}, \nonumber
\end{align*}
i.e. the Euler classical polyatomic gas.

In the case $N=2$ the relativistic moments \eqref{MRel} with $n=0,1$ reduce now to the $15$ moments that generalize the LMR theory to the polyatomic gases \cite{entropy}:
\begin{equation}\label{Rpolino}
\partial_{\alpha} A^\alpha = 0, \quad \partial_{\alpha} A^{\alpha  \beta} = 0, \quad \partial_{\alpha} A^{\alpha  \beta \gamma} = I^{\beta\gamma}
\end{equation}
and the corresponding classical limit is the system:
\begin{align}\label{Cpolino}
& s=0 \,\, \| \quad \partial_t H_0^0 +\partial_i H_0^i = 0,\, \qquad \quad \leftrightarrow   \quad \partial_t F +\partial_i F^i= 0, \quad \,\,   \nonumber\\
&\quad \quad \quad \,\,\,\,\,\, \partial_t H_0^j +\partial_i H_0^{ij} = 0, \qquad \quad \leftrightarrow \quad  \partial_t F^j +\partial_i F^{ij}= 0,    \nonumber \\
&\quad \quad \quad \,\,\,\,\,\, \partial_t H_0^{jk} +\partial_i H_0^{ijk} = J_0^{jk}, \quad \leftrightarrow \quad  \partial_t F^{jk} +\partial_i F^{ijk}= P^{jk},    \\
& s=1 \,\, \| \quad \partial_t H_1^0 +\partial_i H_1^i = 0,\, \qquad \quad \leftrightarrow    \quad  \partial_t G_{ll} +\partial_i G^{lli}= 0, \nonumber \\
&\quad \quad \quad \,\,\,\,\,\, \partial_t H_1^{j} +\partial_i H_1^{ij} = J_1^{j}, \,\, \quad \quad \leftrightarrow \quad  \partial_t G^{llij} +\partial_i G^{lli}= Q^{llj},  \nonumber \\ 
& s=2 \,\, \| \quad \partial_t H_2^0 +\partial_i H_2^i = J_2^0,\,   \nonumber
\end{align}
where the new scalar moment $H_2^0$ and the corresponding flux $H_2^i$ are 
\begin{align}
\left( \begin{array}{l}
  H_2^0\\  H_2^{i}
\end{array}\right)
  = m
 \inta f \,  \left(\xi^2 + 2\frac{\mathcal{I}}{m}\right)^2 
  \left( \begin{array}{c}
  1 \\ \xi_i
\end{array}\right)
   \da , \nonumber
\end{align}
and the production term
\begin{align}
  J_2^0
  = m
 \inta Q \,  \left(\xi^2 + 2\frac{\mathcal{I}}{m}\right)^2 
   \da .\nonumber
\end{align}

\section{Closure of Moments and RET$_{15}$}
Until now, we discussed the choice of the truncated moments to consider and we proved that for a given relativistic system with truncation index $N+1$, there exists a unique classical limit for the moments. The truncated system are both in relativistic and classic regimes, not closed. The closure procedure belongs in RET theory   \cite{ET,RET,book,newbook}.  It is expressed by a hyperbolic system of field equations with local constitutive equations. The closure is obtained at the phenomenological level using the universal principle such as the  entropy principle, the entropy convexity, and the covariance with respect to the  proper group of transformation. Or, at the molecular level, the closure is obtained by using the Maximum Entropy Principle (MEP) introduced in non-equilibrium thermodynamics first by Janes \cite{Janes} and successively developed by M\"uller and Ruggeri that proved as first that the closed system becomes symmetric hyperbolic \cite{ET}.

The closure of polytropic relativistic  Euler fluids \eqref{3.2}  was given first in the paper \cite{Annals} (see also \cite{JMP,ARMA}), while the closure in the case of $15$ fields  (RET$_{15}$)  \eqref{Rpolino} (relativistic case) and \eqref{Cpolino} (classical limit) was respectively  the subject of the  recent papers \cite{entropy} and \cite{IJNM}.

More precisely, in the case $N=2$ the  system \eqref{MRel} becomes
 \begin{equation}\label{Annalis} 
\partial_\alpha A^{\alpha } = 0, \quad \partial_\alpha A^{\alpha \beta} =0, \quad  \partial_\alpha A^{\alpha \beta \gamma} =  I^{  \beta \gamma  }, \qquad \left(\beta,\gamma=0,1,2,3\right). 
\end{equation}
with
\begin{align} \label{relRETpol}
\begin{split}
& A^{\alpha } = {mc}  \inta f  \,  p^\alpha  \, 
\phi(\mathcal{I}) \, d \mathcal{I} \, d \boldsymbol{P}  \, , \\
&A^{\alpha \beta} = c \inta f  \,  p^\alpha p^{\beta}  \, \Big( 1 +  \frac{\II }{m c^2}\Big)\, 
\phi(\mathcal{I}) \, d \mathcal{I} \, d \boldsymbol{P}  \, , \\
& A^{\alpha \beta\gamma } = \frac{c}{m}  \inta  f  \,  p^\alpha p^{\beta}  p^{\gamma}  \, \Big( 1 +  \frac{\II}{mc^2} \Big)^2\, 
\phi(\mathcal{I}) \, d \mathcal{I} \, d \boldsymbol{P}  \, , \\
& I^{ \beta \gamma } = \frac{c}{m} \inta Q  \,  p^{\beta}  p^{\gamma}\, \left(1 +  \frac{\II}{mc^2} \right)^2\, 
\phi(\mathcal{I}) \, d \mathcal{I} \, d \boldsymbol{P}. \\
\end{split}
\end{align}
Recalling the following decomposition of the particle number vector and the energy-momentum tensor in terms of physical variables:
\begin{align*}
A^\alpha =\rho  U^\alpha \, , \quad  A^{\alpha \beta} = \frac{e}{c^2} \,  U^{\alpha } U^\beta + \, \left(p \, + \, \Pi\right)
h^{\alpha \beta} + \frac{1}{c^2} ( U^\alpha  q^\beta +U^\beta  q^\alpha)+   t^{<\alpha \beta>_3} \, ,
\end{align*}
  where $n, \rho = n m, U^\alpha, h^{\alpha\beta},p,e$  are respectively the   particle number, the rest mass density, the four-velocity, the projector tensor $(h^{\alpha\beta}= U^\alpha U^\beta/c^2 - g^{\alpha\beta})$, the pressure, the energy.  Moreover  $g^{\alpha \beta}= \text{diag}(1 \, , \, -1 \, , \, -1 \,, \, -1)$ is the metric tensor,  $\Pi$ is the dynamic pressure, $q^\alpha= -h^\alpha_\mu
U_\nu T^{\mu \nu}$ is the heat flux and $t^{<\alpha \beta>_3} = T^{\mu\nu} \left(h^\alpha_\mu h^\beta_\nu - \frac{1}{3}h^{\alpha\beta}h_{\mu\nu}\right)$ is the deviatoric shear viscous stress tensor.  We also recall the constraints:
\begin{equation*}
U^\alpha U_\alpha = c^2, \quad q^\alpha U_\alpha = 0, \quad t^{<\alpha \beta>_3} U_\alpha = 0, \quad t^{<\alpha}_{\,\,\,\,\,\ \alpha >_3} =0,
\end{equation*}
and we choose as the $15$th variable:
\begin{align*}
\Delta = \frac{4}{c^2} \,  U_\alpha U_\beta U_\gamma \,  \left( A^{\alpha \beta \gamma} \, - \,  A^{\alpha \beta \gamma}_E\right).
\end{align*}
The pressure and the energy 
compatible with the equilibrium distribution function  are \cite{Annals}:
\begin{align}\label{10}
\begin{split}
& p =\frac{ k_B}{m} \, \rho  T \, , \qquad \qquad e=
\rho  c^2   \omega(\gamma), \\
& \text{with } \quad \omega(\gamma)= \frac{\int_0^{+\infty} J_{2,2}^* \, \Big( 1 + \frac{\mathcal{I}}{m c^2} \Big) \, \phi(\mathcal{I})  \, d \, \mathcal{I}}{\int_0^{+\infty} J_{2,1}^* \,  \phi(\mathcal{I})  \, d \, \mathcal{I}},
\end{split}
\end{align}
\begin{equation}\label{8}
\begin{split}
& J_{m,n}^* = J_{m,n} (\gamma^*), \qquad \gamma^* = \gamma \, \Big( 1
+ \frac{\mathcal{I}}{m \, c^2} \Big), \qquad   \gamma = \frac{m \, c^2}{k_B T},
\end{split}
\end{equation}
with $T$ being the temperature and $k_B$ being the Boltzmann constant, and
\begin{equation*}
J_{m,n}(\gamma)= \int_0^{+\infty} e^{-\gamma \cosh s}  \sinh^m s \cosh^n s \, d  s .
\end{equation*}
To close the system \eqref{relRETpol}, we have adopted in \cite{entropy} the MEP which requires to find the distribution function that maximizes the non-equilibrium entropy density:
\begin{equation}\label{Entropy}
h= h^\alpha U_\alpha  \quad \rightarrow \quad \max
\end{equation}
with  the entropy four-vector given by
\begin{equation} \label{fourentropy}
h^\alpha	=  - k_B \, c \,   \inta f
	\ln f p^\alpha \phi(\mathcal{I})  \,  d\mathcal{I}  \,  d\boldsymbol{P},
\end{equation}
under the constraints  that the temporal part $A^\alpha U_\alpha, A^{\alpha\beta}U_\alpha$ and  $A^{\alpha\beta\gamma}U_\alpha$  are prescribed. Proceeding in the usual way as indicated in previous papers of RET (see \cite{RS,Annals}), we obtain:
\begin{align}\label{f15}
\begin{split}
& f= e^{   -1 -  \frac{\chi}{k_B}}   \, , \quad \mbox{with} \\
& \chi = m \, \lambda \, + \,  \lambda_{\mu} \, p^{\mu} \, \left( 1 + \, \frac{\mathcal{I}}{m \, c^2} \right) \, + \,  \frac{1}{m} \, \lambda_{\mu \nu} \, p^{\mu} p^{\nu}  \, \left( 1 + \, \frac{\mathcal{I}}{m \, c^2} \right)^2,
\end{split}
\end{align}
where $\lambda, \lambda_{\mu}, \lambda_{\mu \nu}$ are the Lagrange multipliers.

In the molecular RET approach,  we consider, as usual, the processes near equilibrium.  For this reason, we expand \eqref{f15} around an equilibrium state as follows:
\begin{align*}
\begin{split}
&f \simeq f_E\left(1-\frac{1}{k_B}\tilde{\chi}\right),   \quad \mbox{with} \\
&\tilde{\chi}   = m \, (\lambda - \lambda_E) \, + \,  (\lambda_{\mu}-\lambda_{\mu_E}) \, p^{\mu} \, \left( 1 + \, \frac{\mathcal{I}}{m \, c^2} \right) \, + \,  \frac{1}{m} \, \lambda_{\mu \nu} \, p^{\mu} p^{\nu}  \, \left( 1 + \, \frac{\mathcal{I}}{m \, c^2} \right)^2,
\end{split}
\end{align*}
with
\begin{align*}
\begin{split}
& {\lambda}_E = - \frac{1}{T}\left(g+ c^2\right), \quad {\lambda}_{\mu_E}= \frac{U_\mu}{T},  \quad \lambda_{\mu \nu_E} =0,
\end{split} 
\end{align*}
where $g =\varepsilon +p/\rho -T S $ is the equilibrium chemical potential with $S$ being the equilibrium entropy and $\varepsilon = e/\rho - c^2$.  

In \cite{entropy}, it was proved that choosing as collisional term a variant of BGK model proposed in \cite{BGKPR}  the triple tensor and the production term have necessarily this closed form:
\begin{align*}
\begin{split}
A^{\alpha \beta \gamma} &= \left( \rho\, \theta_{0,2}  + \, \frac{1}{4c^4} \, \Delta \right)  U^{\alpha } U^{\beta} U^\gamma + \left( \rho \, c^2\,\theta_{1,2}  - \, \frac{3}{4c^2}\, \frac{N^\Delta}{D_4} \, \Delta \, - 3 \, \frac{N^\Pi}{D_4} \, \Pi \right)
\,  h^{(\alpha\beta} U^{\gamma)} \\
& + \frac{3}{c^2} \frac{N_3}{D_3}  \, q^{(\alpha} U^\beta U^{\gamma)} + \frac{3}{5} \frac{N_{31}}{D_3} h^{(\alpha
	\beta} q^{\gamma)} + 3  C_5 t^{(<\alpha \beta >_3} U^{\gamma)}
\, ,
\end{split}
\end{align*}
 \begin{align*}
I^{\beta\gamma} = &\frac{1}{\tau}
\Big\{    
- \frac{1}{4c^4  } \Delta  \,  U^\beta U^{\gamma } + \left( \frac{1}{4c^2} \frac{N^\Delta}{D_4}\Delta +\frac{N^\Pi}{D_4} \Pi \right) h^{\beta\gamma} + \\
&\Big( -\frac{2}{c^2 } \frac{N_3}{D_3} +\frac{\theta_{1,3}}{\theta_{1,2}}\frac{1}{c^2 } \Big)q^{(\beta} U^{ \gamma )} -   C_5 t^{<\beta\gamma>_3}
\Big\},
\end{align*}
where all coefficients are explicit functions of 
\begin{align}\label{11b}
\theta_{a,b} = \frac{1}{2a+1}  \begin{pmatrix}
b+1 \\ 2a
\end{pmatrix} \frac{\int_0^{+\infty} J_{2a+2,b+1-2a}^* \, \Big( 1 + \frac{\mathcal{I}}{m c^2} \Big)^b \, \phi(\mathcal{I})  \, d \, \mathcal{I}}{\int_0^{+\infty} J_{2,1}^* \,  \phi(\mathcal{I})  \, d \, \mathcal{I}} \,
\end{align}
that are dimensionless  function only of  $\gamma$ (i.e.  function  of the temperature, see \eqref{8}) and depending  by recursive formula of the unique  function $\omega(\gamma)$ strictly related with  the energy trough the relations \eqref{10}. 

It was also proved in \cite{entropy} that the classical limit of this model coincides with the corresponding classical RET$_{15}$ studied in \cite{IJNM}.

Both the model of relativistic and classical RET$_{15}$ are very complex, and therefore, in principle, it is hard to discuss the qualitative analysis. Nevertheless, we want to prove that they belong to the systems of balance laws with a convex entropy for which exists general theorems of qualitative analysis as we summarize in the next section.

\section{Qualitative Analysis}
The system \eqref{Annalis} belongs to a general quasi-linear system of $N$ balance laws:
\begin{equation}
\partial _{\alpha }\mathbf{F}^{\alpha }(\mathbf{u})=\mathbf{f}(\mathbf{u}),
\label{r1}
\end{equation}%
compatible with an entropy law  
\begin{equation}
\partial _{\alpha }h^{\alpha }(\mathbf{u)}=\Sigma \,(\mathbf{u)},   \qquad \Sigma \geq 0,
\label{e8}
\end{equation}
where $h^\alpha$   and $\Sigma$  are,
respectively, the entropy vector and the entropy production.  
For this kind of systems starting from previous results of Godunov \cite{Godunov}, Friedrichs and Lax \cite{FL} and Boillat \cite{Boillat}, Ruggeri and Strumia proved the following theorem \cite{RS}:
\begin{theorem}[Ruggeri-Strumia]
	\label{CampoPRIN}
	The compatibility between the system of balance laws \eqref{r1} and the supplementary balance law \eqref{e8} with the entropy $h=h^\alpha \xi_\alpha$ being a convex function of $\mathbf{u} \equiv
	\mathbf{F}^\alpha \xi_\alpha$, with $\xi_\alpha$ a congruence time-like, implies the existence of the  "main field" $ \mathbf{u}^\prime $ that satisfies
	\begin{equation*}
	dh^\alpha = \mathbf{u}^\prime \cdot d\mathbf{F}^\alpha, \qquad \Sigma =
	\mathbf{u}^\prime \cdot \mathbf{f}\,\geq 0.  
	\end{equation*}
	If we choose the components of $ \mathbf{u}^\prime $ as field variables, 
	we have
	\begin{equation}
	\mathbf{F}^\alpha = \frac{\partial h^{\prime \alpha}}{\partial \mathbf{u}%
		^\prime},  \label{potenzialis}
	\end{equation}
and 	the 
	original system \textrm{(\ref{r1})} can be rewritten in
	a symmetric form with Hessian matrices:
	\begin{equation}
	\partial_\alpha \left( \frac{\partial h^{\prime \alpha}}{\partial \mathbf{u}%
		^\prime } \right)= \mathbf{f} \quad \iff \quad \frac{\partial^2 h^{\prime
			\alpha}}{\partial \mathbf{u}^\prime \partial \mathbf{u}^\prime}
	\partial_\alpha \mathbf{u}^\prime= \mathbf{f},  \label{simetric}
	\end{equation}
	where $h^{\prime \alpha}$ is the four-potential defined by
	\begin{equation}
	h^{\prime \alpha}= \mathbf{u}^\prime \cdot \mathbf{F}^\alpha - h^\alpha.
	\label{110}
	\end{equation} 
The function
\begin{equation*}
	h^\prime = h^{\prime \alpha} \xi_\alpha =  \mathbf{u}^\prime \cdot \mathbf{u} - h,
\end{equation*}
is the Legendre transformation of $h$ and therefore a convex function of the dual field $\mathbf{u}^\prime$.
\end{theorem}

In the general theory of symmetric  hyperbolic balance laws, it is well-known that the system
(\ref{r1}) has a unique local (in time) smooth solution for smooth
initial data \cite{FL,Kawa,fisher}.
 However, in a general case, even for arbitrarily small and smooth
initial data, there is no global continuation for these smooth solutions,
which may develop singularities, shocks, or blow{up, in a finite time, see for
	instance \cite{maida,dafermos}.

	On the other hand, in many physical examples, thanks to the
	interplay between the source term and the hyperbolicity, there
	exist global smooth solutions for a suitable set of initial data.
 In this context, the following \emph{K-condition} \cite{SK} plays an important role:
		\begin{definition}[K-condition]
		A system (\ref{r1}) satisfies the K-condition if, in 
		the equilibrium manifold,  any right characteristic eigenvectors ${\bf d}$ of (\ref{r1})
		are not in the null space of $\nabla {\bf f}$, where $\nabla\equiv \partial/\partial {\bf u}$:
		\begin{equation}\label{Kcond}
		\left(\nabla {\bf f}\,\, {\bf d}^I\right)_E \neq 0 \quad \forall \, {\bf d}^I, \,\,\,\, I=1,2,\dots N.
		\end{equation}
	\end{definition}
	For dissipative one-dimensional systems (\ref{r1}) satisfying
	the K-condition, it is possible to prove the following global
	existence theorem by Hanouzet and Natalini \cite{nat}:

	\begin{theorem}[Global Existence]  \label{global_e}  
		{Assume that the system} (\ref{r1}) {is strictly dissipative} with a convex entropy
		{and that the
			K-condition  is satisfied. Then there exists $\delta>0$, such that,
			if $\left\Vert \mathbf{u}^{\prime }(x,0)\right\Vert _{2}\leq \delta ,$ there is
			a unique global smooth solution, which verifies}
		\[
		\mathbf{u}^\prime \in C^{0}([0,\infty );H^{2}(R)\cap
		C^{1}([0,\infty );H^{1}(R)).
		\]
	\end{theorem}
	This global existence theorem was generalized to a higher-dimensional case by Yong \cite{wen} and successively by Bianchini, Hanouzet, and Natalini \cite{bnat}.

	Moreover {Ruggeri} and {Serre} \cite{RugSerre} proved that the constant equilibrium 
	state is stable.
 Dafermos showed 
	the existence and long time behavior of spatially periodic BV solutions \cite{Dafermos-2015}. 
	 
	The K-condition is
	only a sufficient condition for the global existence of smooth solutions.
		Lou    and Ruggeri  \cite{Palermo} observed that there indeed exists a weaker K-condition that is a necessary (but unfortunately not sufficient) condition
	for the global existence of smooth solutions.  Instead of the condition that the right eigenvectors are not in the null space of $\nabla {\bf f}$, they posed this condition only on the right eigenvectors corresponding to genuine nonlinear eigenvalues.
	It was proved that the assumptions of the previous theorems are fulfilled in both 
	classical  \cite{Ruget} and relativistic \cite{RugETR,RUGCHO} RET theories of monatomic gases, and also in the theory of mixtures of gases with multi-temperature \cite{SR}.
	
In \cite{entropy}, it was proved that at least in a neighborhood of equilibrium, the entropy \eqref{Entropy}
is a convex function of the field $\mathbf{u} = \mathbf{F}^\alpha U_\alpha$, and the entropy principle \eqref{e8} is satisfied, then we need to prove only the K-condition to satisfy the assumptions of previous theorem.  

For this aim  we first need to calculate the characteristic velocities evaluated in equilibrium. 

\subsection{Characteristic velocities in equilibrium}
We recall that in \cite{Boillat-1997} it was proved that in the theory of moments, the main field coincides with the Lagrange multipliers of MEP (see also \cite{BRrel1,newbook}), and therefore \eqref{MRel} taking into account \eqref{potenzialis} and \eqref{simetric}
can be written 
\begin{equation}\label{1}
\partial_\alpha \left( \frac{\partial \, h'^\alpha}{\partial \, \lambda_A}\right) = I^A \quad , \quad \mbox{or} \quad \frac{\partial^2 \, h'^\alpha}{\partial \, \lambda_A \partial \, \lambda_B} \, \partial_\alpha   \lambda_B = I^A \, ,
\end{equation}
where the multi-index $A$ is used for the Lagrange multipliers in equivalent way of \eqref{F-Mom}:
\begin{equation*} 
\lambda_A = \lambda_{\alpha_1 \alpha_2\dots \alpha_A}, \quad (\lambda \,\, \text{when} \,\, A=0),
\end{equation*}
that in the present case of $15$ moments $A=0,1,2$, i.e. $\mathbf{u}^\prime \equiv (\lambda,\lambda_\alpha,\lambda_{\alpha\beta})$.
	
	As it is well-known, the wave equations associated with the system \eqref{1} can be obtained by the following rule:
	\begin{equation*}
	\partial_\alpha \,\, \rightarrow \,\, \varphi_\alpha\,  \delta\,\ , \qquad I^A \,\, \rightarrow \,\, 0,
	\end{equation*}
with
\begin{align*}
\varphi_\alpha=  \, \frac{V}{c} \, \xi_\alpha + \eta_\alpha \, , 
\end{align*}
where $V$ indicates the characteristic velocity and $\xi_\alpha$ and $ \eta_\alpha$ indicate, respectively, a generic time-like and space-like congruence:  $\xi_\alpha \xi^\alpha=1$, $\xi_\alpha \eta^\alpha=0$, $\eta_\alpha\eta^\alpha=-1$. 
	Therefore we have from \eqref{1}:
\begin{align}\label{4.1}
\varphi_\alpha  \, \frac{\partial^2 \, h'^\alpha}{\partial \, \lambda_A \, \partial \, \lambda_B} \,  \delta  \lambda_B = 0,  
\end{align}
where $\delta  \lambda_B$ are the right eigenvectors associated to the system \eqref{1}.
For the symmetry of $\frac{\partial^2  h'^\alpha}{\partial \lambda_A \, \partial   \lambda_B}$ and    the convexity of $h^\prime = h^{\prime\alpha}\xi_\alpha$ with respect to the main field, the quadratic form \footnote{We recall in Mathematical community the entropy is the physical entropy changed by sign and therefore we use still  the terms convexity where in reality our function is  concave.}
\begin{equation*}
  \frac{\partial^2  h^\prime}{\partial   \lambda_A \, \partial   \lambda_B} \, \delta  \lambda_A \, \delta  \lambda_B 
\end{equation*}
   is negative definite $\forall$ time-like 4-vector $\xi_\alpha$, we can deduce that all the eigenvalues $V$ are real and the equations \eqref{4.1} gives a basis of eigenvectors $\delta  \lambda_B$; in other words, our field equations, according with Theorem \ref{CampoPRIN},  is symmetric hyperbolic. \\
We have also that the characteristic velocities $ V$ don' t exceed the light speed, i.e., $V^2 \leq c^2$, thanks to theorems proved in \cite{BRrel1} and  in the  Appendix A of \cite{Hyp}. \\
  
For the effective evaluation of the wave velocities, we use the same strategy used in a similar problem  in \cite{Hyp}. More precisely, by using the definition of the 4-potential \eqref{110}, the expressions \eqref{relRETpol},   $f$ given in \eqref{f15} and the entropy vector \eqref{fourentropy}, we obtain:
\begin{align*}
h'^\alpha  =   - \, k_B c \,  \inta  \, e^{- \, 1 \, - \, \frac{\chi}{k_B}} \, p^\alpha  \phi(\mathcal{I})  \,   d \,  \mathcal{I}  \, d \boldsymbol{P} \, ,
\end{align*}
then
\begin{align*}
\frac{\partial \, h'^\alpha }{\partial \, \lambda_A} =    c \,  \inta  \, e^{- \, 1 \, - \, \frac{\chi}{k_B}} \, \frac{\partial \, \chi }{\partial \, \lambda_A}  \, p^\alpha  \phi(\mathcal{I})  \,   d \,  \mathcal{I}  \, d \boldsymbol{P} \, .
\end{align*}
Since from \eqref{f15} $\chi$ is linear in the Lagrange multipliers, $\frac{\partial  \chi }{\partial     \lambda_A}$ does not depend on $\lambda_B$, it follows 
\begin{align*}
\frac{\partial^2 \, h'^\alpha }{\partial \, \lambda_B \, \partial \, \lambda_A} = - \,   \frac{c}{k_B} \,  \inta  \, e^{- \, 1 \, - \, \frac{\chi}{k_B}} \, \frac{\partial \, \chi }{\partial \, \lambda_B}  \, \frac{\partial \, \chi }{\partial \, \lambda_A}  \, p^\alpha  \phi(\mathcal{I})  \,   d \,  \mathcal{I}  \, d \boldsymbol{P} \, .
\end{align*}
By using these results,  we can consider the quadratic form 
\begin{align*}
 \delta K=  - \,   \frac{c}{k_B} \, \varphi_\alpha \,  \inta  \, e^{- \, 1 \, - \, \frac{\chi}{k_B}} \, (\delta \, \chi \, )^2 \, p^\alpha  \phi(\mathcal{I})  \,   d \,  \mathcal{I}  \, d \boldsymbol{P} \, , 
\end{align*}
and see that the equations \eqref{4.1} for the wave velocities are equivalent to say that the derivatives of $\delta K$ with respect to $ \delta \lambda _A$ are zero.\\
As  the closure was obtained only near equilibrium, we rewrite 
$ \delta K$  as \begin{align*}
\begin{split}
&  \delta K=   \varphi_\alpha  \, \left[ \frac{\partial^2 \, h'^\alpha }{\partial \, \lambda^2} \, \left( \delta  \lambda \right)^2  \, + 2 \, \frac{\partial^2 \, h'^\alpha }{\partial \, \lambda \, \partial \, \lambda_\mu} \,  \delta  \lambda \,  \delta  \lambda_\mu +  2 \, \frac{\partial^2 \, h'^\alpha }{\partial \, \lambda \, \partial \, \lambda_{\mu \nu}} \,  \delta  \lambda \,  \delta  \lambda_{\mu \nu} + \right. \\
& \left. +   \, \frac{\partial^2 \, h'^\alpha }{\partial \, \lambda_\beta \, \partial \, \lambda_\mu} \,  \delta  \lambda_\beta \,  \delta  \lambda_\mu +  2 \, \frac{\partial^2 \, h'^\alpha }{\partial \, \lambda_\beta \, \partial \, \lambda_{\mu \nu}} \,  \delta  \lambda_\beta \,  \delta  \lambda_{\mu \nu} + \frac{\partial^2 \, h'^\alpha }{\partial \, \lambda_{\beta \gamma} \, \partial \, \lambda_{\mu \nu}} \,  \delta  \lambda_{\beta \gamma}  \,  \delta  \lambda_{\mu \nu} \right]
\, .
\end{split}
\end{align*}
By calculating the coefficients at equilibrium, it becomes 
\begin{align*}
\begin{split}
& \delta K_E=   - \, \frac{m}{k_B} \, \varphi_\alpha  \, \left[ A_E^\alpha  \, \left( \delta  \lambda \right)^2  \, + 2 \, A_E^{\alpha \mu}  \,  \delta  \lambda \,  \delta  \lambda_\mu +   2 \, A_{E}^{\alpha  \mu \nu}   \,  \delta  \lambda \,  \delta  \lambda_{\mu \nu} \, + \right. \\
& \quad \quad  +  \, A_{E}^{\alpha \beta \delta} \,  \delta  \lambda_\beta \,  \delta  \lambda_\delta + \left. 2 \, A_{E}^{\alpha \beta \mu \nu} \,  \delta  \lambda_\beta \,  \delta  \lambda_{\mu \nu} +  \, A_{E}^{\alpha \beta \gamma \mu \nu}  \,  \delta  \lambda_{\beta \gamma}  \,  \delta  \lambda_{\mu \nu} \right]
\, ,
\end{split}
\end{align*}
where the explicit  expressions of the tensors in the right hand side in terms of the $\vartheta_{a,b}$ \eqref{11b}  are  reported in \cite{entropy}. For the sake of simplicity, we calculate also the coefficients of the differentials in the reference frame where $U^\alpha$ and $\varphi^\alpha$ have the components  $U^\alpha \equiv (c \, , \, 0  \, , \, 0 \, , \, 0)$ and $\varphi_\alpha \equiv (\varphi_0 \, , \, \varphi_1  \, , \, 0 \, , \, 0)$; in any case, we can at the end express again all the results in covariant form replacing $\varphi_0$ and $\left( \varphi_1 \right)^2$ with $\varphi_0 = \frac{1}{c} \, \varphi^\alpha U_\alpha$ and $\left( \varphi_1 \right)^2=  \varphi_\alpha \varphi_\beta h^{\alpha \beta}$. 

After having calculated $ \delta K_E$, we note that a first eigenvalue is 
\begin{align}\label{eig1}
\varphi_0=0 \, , \quad \mbox{i.e., } \quad V = -c \, \frac{U^\alpha \eta_\alpha}{U^\gamma \xi_\gamma} \, , \quad \mbox{i.e., } \quad V =0 \, , 
\end{align}
where the last expression holds when $\xi_\gamma =   U_\gamma/c$. 

We can observe that   $\varphi_1 \neq 0$ under the hypothesis that the 2 time-like vectors $\xi_\gamma$ and  $U_\gamma$ are oriented both towards the future or both towards the past. After that, if $\varphi_0=0$, the derivatives of $ \delta K_E$ with respect to $ \delta \lambda _A$ give a system whose solution is  $ \delta \lambda _1=0$, $ \delta \lambda _{01}=0$, $ \delta \lambda _{12}=0$, $ \delta \lambda _{13}=0$ and the remaining  unknowns are linked only by 
\begin{align}\label{eig1bis}
\begin{split}
& 6 \, \vartheta_{1,1} \, \delta  \lambda \, + \, 2 \, \vartheta_{1,2} \, c \,  \delta  \lambda_{0} \, + \,   \vartheta_{1,3} \, c^2 \,  \delta  \lambda_{00} \, + \, 
2 \, \vartheta_{2,3} \, c^2 \, 
\left( 3 \, \delta  \lambda_{11} \, +  \, \delta  \lambda_{22} \,  + \, \delta  \lambda_{33} \right) =0 \, , \\
& 10 \, \vartheta_{1,2} \, \delta  \lambda \, + \, 5 \, \vartheta_{1,3} \, c \,  \delta  \lambda_{0} \, + \, 3 \, \vartheta_{1,4} \, c^2 \,  \delta  \lambda_{00} \, + \,   2 \, \vartheta_{2,4} \, c^2 \, 
\left( 3 \, \delta  \lambda_{11} \, +  \, \delta  \lambda_{22} \,  + \, \delta  \lambda_{33} \right) =0  \, , \\
& 5 \, \vartheta_{2,3} \, \delta  \lambda_2 \, + \, 2 \, \vartheta_{2,4} \, c \,  \delta  \lambda_{20} =0 \, , \quad 
5 \, \vartheta_{2,3} \, \delta  \lambda_3 \, + \, 2 \, \vartheta_{2,4} \, c \,  \delta  \lambda_{30} =0 \, .
\end{split}
\end{align}
Therefore, we have 7 free unknowns and the eigenvalue \eqref{eig1} has multiplicity 7. 

For the research of other eigenvalues we have  $\varphi_0 \neq 0$ and we can consider the quadratic form $ - \, \frac{k_B}{m \, \rho \, c \, \varphi_0} \,  \delta K_E $. By defining 

\noindent
$ \delta \lambda  = X_1$,  $c \, \delta  \lambda_{0} = X_2$, $c \, \delta  \lambda_{1} = X_3$, $c^2 \, \delta  \lambda_{00} = X_4$, $c^2 \, \delta  \lambda_{01} = X_5$, $c^2 \, \delta  \lambda_{11} = X_6$, $c^2 \,\left( \delta  \lambda_{22} \, + \, \delta  \lambda_{33} \right) = X_7$,  $c \, \delta  \lambda_{2} = Y_1$,  $c^2 \, \delta  \lambda_{20} = Y_2$,  $c^2 \, \delta  \lambda_{12} = Y_3$, 
$c \, \delta  \lambda_{3} = Z_1$, \\ 
$c^2 \, \delta  \lambda_{30} = Z_2$,  $c^2 \, \delta  \lambda_{13} = Z_3$, $c^2 \delta  \lambda_{23} = Y_4$,
$c^2 \left( \delta  \lambda_{22} \, - \, \delta  \lambda_{33 }\right) = Z_4$,

\noindent
we have
\begin{align*}
- \, \frac{k_B}{m \, \rho \, c \, \varphi_0} \,  \delta K_E  = \sum_{h,k=1}^{7} \, a_{hk} \, X_h \, X_k \, + \, \sum_{h=1}^{3} \, b_{hk} \, Y_h \, Y_k \, + \,  \sum_{h=1}^{3} \, b_{hk} \, Z_h \, Z_k \, + \\ 
+ \, \frac{4}{15}  \, \vartheta_{2,4} \, \left( Y_4 \right)^2 \, + \, \frac{1}{15}  \, \vartheta_{2,4} \, \left( Z_4 \right)^2 \, ,
\end{align*}
with
\begin{align*}
\begin{split}
& a_{11} = \vartheta_{0,0}  \, , \quad a_{12} = \vartheta_{0,1} \, , \quad  a_{13} =  \vartheta_{1,1} \, \frac{\varphi_1 }{\varphi_0}  \, , \quad  a_{14} =  \vartheta_{0,2}  \, , \quad  a_{15} = \frac{2}{3}  \,  \vartheta_{1,2} \, \frac{\varphi_1 }{\varphi_0} \, , \\
& a_{16} = \frac{1}{3}  \, \vartheta_{1,2} \, , \quad a_{17} = a_{16} \, , \quad a_{22} =   \vartheta_{0,2} \, , \quad  a_{23} = \frac{1}{3}  \, \vartheta_{1,2} \, \frac{\varphi_1 }{\varphi_0}  \, , \\  
& a_{24} =   \vartheta_{0,3}  \, , \quad  a_{25} = \frac{1}{3}   \, \vartheta_{1,3} \, \frac{\varphi_1 }{\varphi_0} \, , \quad a_{26} = \frac{1}{6} \, \vartheta_{1,3} \, , \quad  a_{27} = a_{26} \, , \quad  a_{33} = \frac{1}{3}  \, \vartheta_{1,2} \, , \\
&  a_{34} = \frac{1}{6}  \, \vartheta_{1,3} \, \frac{\varphi_1 }{\varphi_0} \, , \quad a_{35} = \frac{1}{3}  \, \vartheta_{1,3} \, , \quad  a_{36} = \vartheta_{2,3} \, \frac{\varphi_1 }{\varphi_0}  \, ,  \quad a_{37} = \frac{1}{3}  \, \vartheta_{2,3} \, \frac{\varphi_1 }{\varphi_0} \, , \\
& a_{44} =  \vartheta_{0,4}  \, , \quad  a_{45} = \frac{1}{5}  \, \vartheta_{1,4} \, \frac{\varphi_1 }{\varphi_0} \, , \quad  a_{46} = \frac{1}{10}  \, \vartheta_{1,4} \, , \quad  a_{47} =  a_{46} \, , \quad  a_{55} = \frac{2}{5} \, \vartheta_{1,4} \, , \\  
& a_{56} = \frac{2}{5} \, \vartheta_{2,4} \, \frac{\varphi_1 }{\varphi_0} \, \quad  a_{57} = \frac{2}{15}  \, \vartheta_{2,4} \, \frac{\varphi_1 }{\varphi_0} \, , \quad  a_{66} = \frac{1}{5}  \, \vartheta_{2,4} \, , \quad  a_{67} = \frac{1}{15}  \, \vartheta_{2,4} \, , \\
& a_{77} = \frac{2}{15}   \, 
\vartheta_{2,4} \, , 
\end{split}
\end{align*}
\begin{align*}
\begin{split}
& b_{11} = \frac{1}{3}  \,\vartheta_{1,2} \, , \quad   b_{12} = \frac{1}{3}  \, \vartheta_{1,3} \, , \quad  b_{13} = \frac{2}{3}  \, \vartheta_{2,3} \, \frac{\varphi_1 }{\varphi_0} \, , \\
& b_{22} = \frac{2}{5}  \,\vartheta_{1,4} \, , \quad   b_{23} = \frac{4}{15}  \, \vartheta_{2,4} \, \frac{\varphi_1 }{\varphi_0} \, , \quad  b_{33} = \frac{4}{15}  \, \vartheta_{2,4}   \, . 
\end{split}
\end{align*}
From these results it follows that the equations to determine eigenvalues and eigenvectors are
\begin{align}\label{eig2}
\sum_{k=1}^{7} \, a_{hk} \, X_k =0 \, , \quad    \sum_{k=1}^{3} \, b_{hk} \, Y_k =0 \, , \quad   \sum_{k=1}^{3} \, b_{hk} \, Z_k =0 \, , \quad Y_4=0 \,  , \quad Z_4=0 \, .  
\end{align}
Here eqs. \eqref{eig2}$_{4,5}$ are $Y_4=0$, $Z_4=0$. The equations  \eqref{eig2}$_{2,3}$ shows that 2 eigenvalues with multiplicity 2 are the solution of 
\begin{align}\label{eig3}
\det \left\|
\begin{matrix}
\frac{1}{3} \, \vartheta_{1,2} & \frac{1}{3} \, \vartheta_{1,3}  & \frac{2}{3} \, \vartheta_{2,3} \, \frac{\varphi_1 }{\varphi_0} \\
&& \\
\frac{1}{3} \, \vartheta_{1,3} & \frac{2}{5} \, \vartheta_{1,4}  & \frac{4}{15} \, \vartheta_{2,4} \, \frac{\varphi_1 }{\varphi_0} \\
&& \\
\frac{2}{3} \, \vartheta_{2,3} \, \frac{\varphi_1 }{\varphi_0} & \frac{4}{15} \, \vartheta_{2,4} \, \frac{\varphi_1 }{\varphi_0}   & \frac{4}{15} \, \vartheta_{2,4} 
\end{matrix} 
\right\| = 0 \, ,
\end{align}
that is, 
\begin{align*}
\det \left\| 
\begin{matrix}
\vartheta_{1,2} & \vartheta_{1,3}  & 2 \, \vartheta_{2,3}  \\
&& \\
\vartheta_{1,3} & \frac{6}{5} \, \vartheta_{1,4}  & \frac{4}{5} \, \vartheta_{2,4}  \\
&& \\
2 \, \vartheta_{2,3}  & \frac{4}{5} \, \vartheta_{2,4}   & 0
\end{matrix} 
\right\| \, \left( \frac{\varphi_1 }{\varphi_0} \right)^2 \, + \, \frac{4}{5} \, \vartheta_{2,4}  \, \det \left\| \begin{matrix}
\vartheta_{1,2} & \vartheta_{1,3}    \\
& \\
\vartheta_{1,3} & \frac{6}{5} \, \vartheta_{1,4} 
\end{matrix} \right\| 
= 0 \, ,
\end{align*}

The eigenvalues different from those in \eqref{eig1}, \eqref{eig3} are given by \eqref{eig2}$_{1}$ in the unknowns $X_k$, that is the determinant of the matrix $a_{hk}$ must be zero, i.e., 
\begin{align}\label{eig5}
\det \left\| \begin{matrix}
\vartheta_{0,0} & \vartheta_{0,1} & \vartheta_{1,1} \, \frac{\varphi_1 }{\varphi_0} &  \vartheta_{0,2} & \frac{2}{3}   \, \vartheta_{1,2} \, \frac{\varphi_1 }{\varphi_0} & \frac{1}{3}  \, \vartheta_{1,2} & \frac{1}{3} \, \vartheta_{1,2} \\
&&&&&& \\
\vartheta_{0,1} & \vartheta_{0,2} & \frac{1}{3}  \, \vartheta_{1,2} \, \frac{\varphi_1 }{\varphi_0} &  \vartheta_{0,3} & \frac{1}{3}   \, \vartheta_{1,3} \, \frac{\varphi_1 }{\varphi_0} &  \frac{1}{6} \, \vartheta_{1,3} &  \frac{1}{6}\, \vartheta_{1,3} \\
&&&&&& \\
\vartheta_{1,1} \, \frac{\varphi_1 }{\varphi_0} & \frac{1}{3}  \, \vartheta_{1,2} \, \frac{\varphi_1 }{\varphi_0} & \frac{1}{3}  \, \vartheta_{1,2} & \frac{1}{6}  \, \vartheta_{1,3} \, \frac{\varphi_1 }{\varphi_0} & \frac{1}{3}  \, \vartheta_{1,3} &   \vartheta_{2,3} \, \frac{\varphi_1 }{\varphi_0} & \frac{1}{3}  \,  \vartheta_{2,3} \, \frac{\varphi_1 }{\varphi_0} \\
&&&&&&  \\
\vartheta_{0,2} & \vartheta_{0,3} & \frac{1}{6}  \, \vartheta_{1,3} \, \frac{\varphi_1 }{\varphi_0} & \vartheta_{0,4} & \frac{1}{5}  \, \vartheta_{1,4} \, \frac{\varphi_1 }{\varphi_0} & \frac{1}{10} \, \vartheta_{1,4} & \frac{1}{10} \, \vartheta_{1,4} \\
&&&&&& \\
\frac{2}{3}   \, \vartheta_{1,2} \, \frac{\varphi_1 }{\varphi_0} & \frac{1}{3}  \, \vartheta_{1,3} \, \frac{\varphi_1 }{\varphi_0} & \frac{1}{3}  \, \vartheta_{1,3} & \frac{1}{5}  \, \vartheta_{1,4} \, \frac{\varphi_1 }{\varphi_0} & \frac{2}{5}  \, \vartheta_{1,4} & \frac{2}{5}  \, \vartheta_{2,4} \, \frac{\varphi_1 }{\varphi_0} & \frac{2}{15}  \, \vartheta_{2,4} \, \frac{\varphi_1 }{\varphi_0}  \\
&&&&&&  \\
\frac{1}{3}   \, \vartheta_{1,2} & \frac{1}{6}\, \vartheta_{1,3} & \vartheta_{2,3} \, \frac{\varphi_1 }{\varphi_0} & \frac{1}{10}   \, \vartheta_{1,4} & \frac{2}{5}  \, \vartheta_{2,4} \, \frac{\varphi_1 }{\varphi_0} & \frac{1}{5}  \, \vartheta_{2,4} & \frac{1}{15}  \, \vartheta_{2,4} \\
&&&&&& \\
\frac{1}{3}  \, \vartheta_{1,2} & \frac{1}{6} \, \vartheta_{1,3} & \frac{1}{3}  \, \vartheta_{2,3} \, \frac{\varphi_1 }{\varphi_0} & \frac{1}{10}  \, \vartheta_{1,4} & \frac{2}{15} \, \vartheta_{2,4} \, \frac{\varphi_1 }{\varphi_0} & \frac{1}{15} \, \vartheta_{2,4} &   \frac{2}{15} \vartheta_{2,4}  
\end{matrix} \right\| = 0 . \nonumber \\
\nonumber \\
\end{align}
It is easy to prove that this equation depends on $\frac{\varphi_1 }{\varphi_0}$ only through $\left( \frac{\varphi_1 }{\varphi_0} \right)^2= \frac{h^{\alpha \beta} \varphi_\alpha \varphi_\beta}{\left( U^\gamma \varphi_\gamma \right)^2 } \, c^2$ (which is equal to $\left( \frac{c}{V} \right)^2$ if $U^\alpha = c \, \xi^\alpha$) and it is a second degree equation in $\left( \frac{\varphi_1 }{\varphi_0} \right)^2$. 	
So it gives 4 independent eigenvectors; other 7 come from \eqref{eig1}, other 4 come from \eqref{eig3}. The total is 15, as expected. 

\begin{figure}[h!] 
	\centering
	\includegraphics[width=80mm]{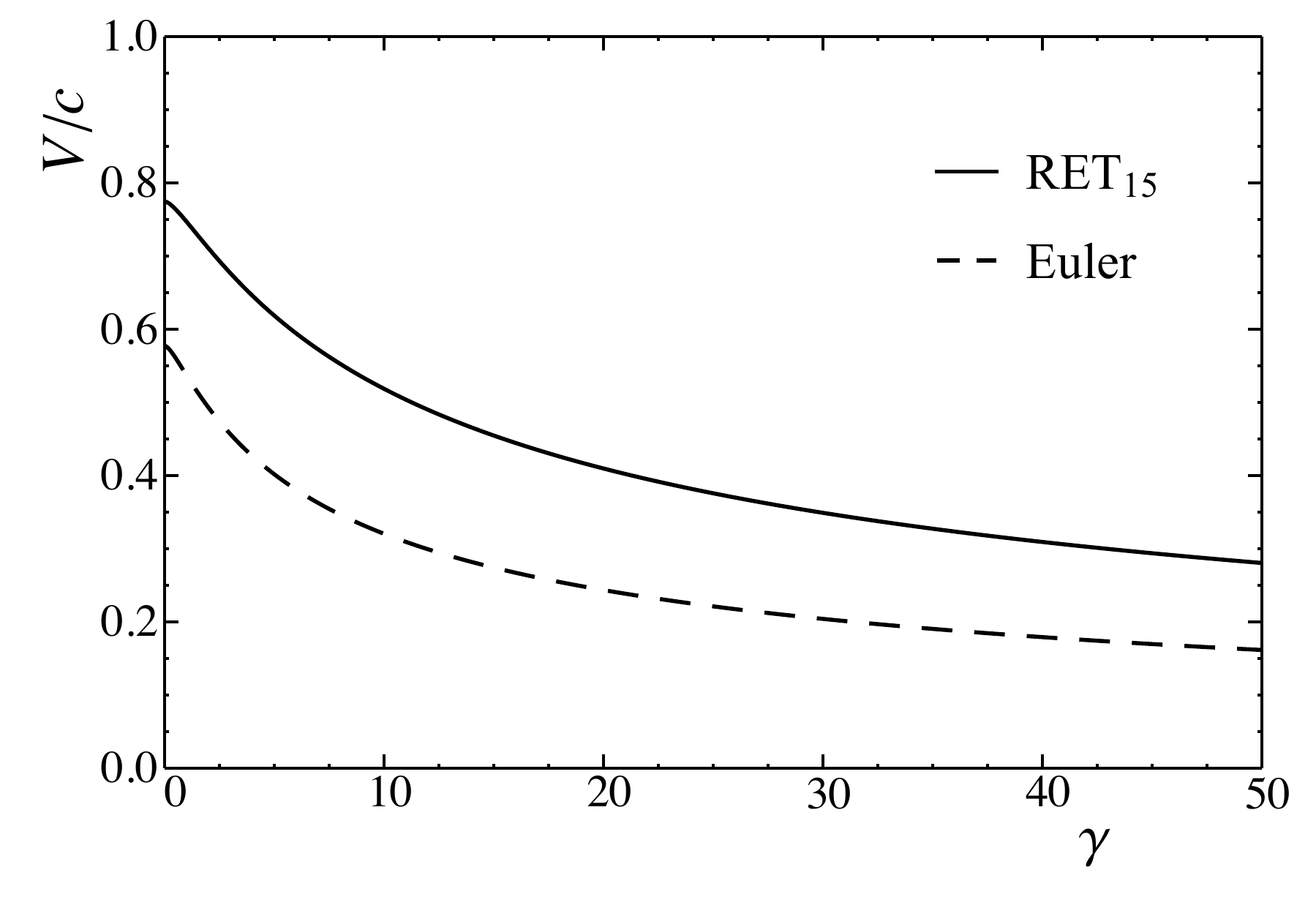}
	\caption{Dependence of maximum  $V/c$ on $\gamma$ for a diatomic gas. The solid line indicates the value of RET$_{15}$  and the dashed line indicates the one of Euler system \cite{ARMA}.}
	\label{CS}
\end{figure}

\bigskip

As example we can use the expressions of $\vartheta_{a,b}$ \eqref{11b}  given   in \cite{entropy} in the case of diatomic gases for which the expression of energy $e$ given in \eqref{10} is explicit because $\omega(\gamma)$ can be written in terms of ratio of modified Bessel functions \cite{ARMA}
\begin{equation*}
	\omega(\gamma)= \frac{K_0(\gamma)}{K_1(\gamma)} +\frac{3}{\gamma}.
\end{equation*} 
As consequence it is easy to plot the maximum characteristic velocity in the rest frame as function of $\gamma$ (see Fig. \ref{CS}). 
According with general results for which increasing the number of moments increases the maximum characteristic velocity \cite{Boillat-1997,BRrel1} and 
since the relativistic Euler system is a\emph{ principal subsystem} of RET$_{15}$ by the definition given in \cite{BoillatRuggeriARMA}, the sub-characteristic conditions hold and  the maximum characteristic velocity of RET$_{15}$ that is obtained from  \eqref{eig5} is larger than the one of Euler system which is studied in \cite{ARMA} as evident in Fig.  \ref{CS}.

\subsection{K-Condition}

As was proved in \cite{Palermo}, 
	the K-condition \eqref{Kcond} is equivalent to $\delta {\mathbf{f}} \neq 0$ 
 for any characteristic velocities in  equilibrium. In the present case this is equivalent to prove that
$\delta  I^{\beta \gamma} \neq 0$   at equilibrium. 
To prove that it holds in our case, let us suppose by absurd that it does't hold, i.e., that there is at least a characteristic velocity with $\delta  I^{\beta \gamma} = 0$. 
Now $\delta \, I^{\beta \gamma} = 0$  with the coefficients of the differentials of the independent variables calculated at equilibrium, is equivalent to $ \delta \lambda _{\beta \gamma} = 0$ (because the quadratic form $\Sigma$ is positive defined and, consequently, $I^{\beta \gamma}$ is invertible in $\lambda_{\beta \gamma}$). 

If the eigenvalue under consideration is $\varphi_\alpha U^\alpha=0$, this means that, jointly with \eqref{eig1bis}, its expression calculated in $ \delta \lambda _{\beta \gamma} = 0$ holds, i.e.,  
\begin{align*}
6 \, \vartheta_{1,1} \, \delta  \lambda \, + \, 4 \, \vartheta_{1,2} \, c \,  \delta  \lambda_{0}  =0, \, \,    10 \, \vartheta_{1,2} \, \delta  \lambda \, + \, 5 \, \vartheta_{1,3} \, c \,  \delta  \lambda_{0}   =0, \,\, \delta  \lambda_2  =0,  \, \,   \delta  \lambda_3 =0.
\end{align*}
This implies $ \delta \lambda _0=0$. Jointly with \eqref{eig1bis} and with the other results written before, we obtain $ \delta \lambda =0$, $ \delta \lambda _\beta=0$,  $ \delta \lambda _{\beta \gamma} = 0$. This absurd shows that the K-condition is satisfied for the eigenvalue  $\varphi_\alpha U^\alpha=0$. 

If the eigenvalue under consideration is one of those in \eqref{eig3}, the absurd hypothesis means that, jointly with \eqref{eig2},  its expression calculated in $ \delta \lambda _{\beta \gamma} = 0$ holds, i.e.,  
\begin{align}\label{eige6}
\sum_{k=1}^{3} \, a_{hk} \, X_k =0 \, , \quad     b_{h1} \, Y_1 =0 \, , \quad    b_{h1} \, Z_1 =0  \, . 
\end{align}
The last two of these relations give $ \delta \lambda _2 = 0 $, $ \delta \lambda _3 = 0 $, while the first one gives ( with the results written before \eqref{eig3}) $ \delta \lambda  = 0 $, $ \delta \lambda _0 = 0 $, $ \delta \lambda _1 = 0 $. This implies  $ \delta \lambda =0$, $ \delta \lambda _\beta=0$,  $ \delta \lambda _{\beta \gamma} = 0$. This absurd result shows that the K-condition is satisfied for the eigenvalues \eqref{eig3}. 

It remains to prove that it also for the eigenvalues which are solutions of \eqref{eig5}. In this case  the absurd hypothesis means that, jointly with \eqref{eig2}, also its expression calculated in $ \delta \lambda _{\beta \gamma} = 0$ holds, i.e.,  \eqref{eige6}. 
The last two of these relations give $ \delta \lambda _2 = 0 $, $ \delta \lambda _3 = 0 $, while the first one says that
\begin{align}\label{eige7}
\left( \begin{matrix}
\vartheta_{0,0} & \vartheta_{0,1} & \vartheta_{1,1} \, \frac{\varphi_1 }{\varphi_0} \\
&& \\
\vartheta_{0,1} & \vartheta_{0,2} & \frac{1}{3}  \, \vartheta_{1,2} \, \frac{\varphi_1 }{\varphi_0}\\
&& \\
\vartheta_{1,1} \, \frac{\varphi_1 }{\varphi_0} & \frac{1}{3}  \, \vartheta_{1,2} \, \frac{\varphi_1 }{\varphi_0} & \frac{1}{3}  \, \vartheta_{1,2} \\
&&  \\
\vartheta_{0,2} & \vartheta_{0,3} & \frac{1}{6}  \, \vartheta_{1,3} \, \frac{\varphi_1 }{\varphi_0} \\
&& \\
\frac{2}{3}   \, \vartheta_{1,2} \, \frac{\varphi_1 }{\varphi_0} & \frac{1}{3}  \, \vartheta_{1,3} \, \frac{\varphi_1 }{\varphi_0} & \frac{1}{3}  \, \vartheta_{1,3} \\
&&  \\
\frac{1}{3}   \, \vartheta_{1,2} & \frac{1}{6}\, \vartheta_{1,3} & \vartheta_{2,3} \, \frac{\varphi_1 }{\varphi_0}  \\
&& \\
\frac{1}{3}  \, \vartheta_{1,2} & \frac{1}{6} \, \vartheta_{1,3} & \frac{1}{3}  \, \vartheta_{2,3} \, \frac{\varphi_1 }{\varphi_0}
\end{matrix} \right) 
\left(\begin{matrix}
 \delta \lambda  \\
c \, \delta  \lambda_0 \\
c \, \delta  \lambda_1
\end{matrix} \right) = \left(\begin{matrix}
0 \\
0 \\
0
\end{matrix} \right) . 
\end{align}
This implies that all the 35 third order minors of the matrix in the left hand side must be zero for the same value of the unknown $\frac{\varphi_1 }{\varphi_0}$. In particular, we may consider eqs. \eqref{eige7}$_{1,2,4}$ (or \eqref{eige7}$_{1,2,6}$, or \eqref{eige7}$_{1,2,7}$, or \eqref{eige7}$_{1,4,6}$, or \eqref{eige7}$_{1,4,7}$, or \eqref{eige7}$_{2,4,6}$, or \eqref{eige7}$_{2,4,7}$, or \eqref{eige7}$_{4,6,7}$) and obtain the result $\frac{\varphi_1 }{\varphi_0}=0$ which is absurd because we have said that $\varphi_1 \neq 0$ and, in any case, the system \eqref{eige7} would give the absurd result 
$ \delta \lambda  =0$, $ \delta \lambda _0 =0$, $ \delta \lambda _1 =0$.

\bigskip

Similar calculations can be done for the classical limit system \eqref{Cpolino}  (see \cite{IJNM}) and it is possible to prove that also in classical case the K-condition is satisfied.

\bigskip

\emph{Therefore we can conclude that both the solutions of the relativistic and classical systems satisfy the theorems before stated, and as a consequence, global smooth solutions exist provided initial data are sufficiently small and not far away from an equilibrium state.}

\bigskip

\bmhead{Acknowledgments}

	The work has been partially supported by JSPS KAKENHI Grant Numbers JP18K13471 (TA),   by the Italian MIUR through the PRIN2017
	project "Multiscale phenomena in Continuum Mechanics:
	singular limits, off-equilibrium and transitions" Project Number:
	2017YBKNCE (SP) and GNFM/INdAM (MCC, SP and TR).

\section*{Declarations}
\begin{itemize}
\item Funding: JSPS KAKENHI Grant Numbers JP18K13471; 
PRIN2017 Number: 2017YBKNCE.
\item Conflict of interest/Competing interests: The authors declare no conflict of interest.
\item  Availability of data and materials: Not applicable.
\end{itemize}

\end{document}